\documentclass[10pt,conference]{IEEEtran}
\IEEEoverridecommandlockouts
\usepackage{cite}
\usepackage{amsmath,amssymb,amsfonts}
\usepackage{algorithmic}
\usepackage{graphicx}
\usepackage{textcomp}
\usepackage{xcolor}
\usepackage{soul}
\def\BibTeX{{\rm B\kern-.05em{\sc i\kern-.025em b}\kern-.08em
    T\kern-.1667em\lower.7ex\hbox{E}\kern-.125emX}}

\pdfoutput=1 

\usepackage{amsmath}
\usepackage{amsthm}
\usepackage{enumitem}
\setlist{itemsep=-0.0ex}
\usepackage{graphicx}
\usepackage{physics}
\usepackage[ruled,lined,linesnumbered]{algorithm2e}

\usepackage{xcolor}
\usepackage{tikz}
\usetikzlibrary{quantikz}
\usepackage{float}
\usepackage{hyperref}
\usepackage{cleveref}
\usepackage{url}

\newcommand{\cond}{-10pt}

\tikzset{
unfilled/.style = {circle, draw = black, minimum size = 0.5cm}
}
\newtheorem{theorem}{Theorem}
\crefalias{theorem}{theorem}
\theoremstyle{definition}

\crefalias{problem}{problem}
\newtheorem{definition}[theorem]{Definition}
\crefalias{definition}{definition}
\newtheorem{example}[theorem]{Example}
\crefalias{example}{example}
\newtheorem{procedure}[theorem]{Procedure}
\crefalias{procedure}{procedure}
\newtheorem{corollary}[theorem]{Corollary}
\crefalias{corollary}{corollary}

\newcommand{\bp}[1] {{\left( #1 \right)}}

\newcommand{\snote}[1]{\textcolor{blue}{[*** Jamie: #1 ***]}} 

\newcommand{\jnote}[1]{\textcolor{red}{[*** Jason: #1 ***]}}

\newcommand{\nnote}[1]{\textcolor{brown}{[*** Nic: #1 ***]}} 
\newcommand{\tnote}[1]{\textcolor{green!70!black}{[*** Travis: #1 ***]}}

\newtheorem*{remark}{Remark}
\begin{document}

\title{Masking Countermeasures Against Side-Channel Attacks on Quantum Computers
\thanks{This research was funded in part by the Commonwealth Cyber Initiative (CCI), an investment in the advancement of cyber R\&D, innovation, and workforce development. For more information about CCI, visit \url{www.cyberinitiative.org}.}
}

\author{ 
\IEEEauthorblockN{Jason T.~LeGrow\IEEEauthorrefmark{1}\IEEEauthorrefmark{3},
Travis Morrison\IEEEauthorrefmark{1},
Jamie Sikora\IEEEauthorrefmark{2}\IEEEauthorrefmark{3},
Nic Swanson\IEEEauthorrefmark{1}
}
\IEEEauthorblockA{
\IEEEauthorrefmark{1} Department of Mathematics \\
\IEEEauthorrefmark{2} Department of Computer Science \\
\IEEEauthorrefmark{3} Center for Quantum Information Science and Engineering \\
Virginia Tech, \\
Blacksburg, Virginia, USA\\
Email: \(\mathtt{\{jlegrow,tmo,sikora,nicswanson\}@vt.edu}\)
}
}

\IEEEpubid{\makebox[\columnwidth]{979-8-3315-4137-8~\copyright2024 IEEE \hfill}
\hspace{\columnsep}\makebox[\columnwidth]{ }}

\maketitle

\begin{abstract}
We propose a modification to the transpiler of a quantum computer to safeguard against side-channel attacks aimed at learning information about a quantum circuit. We demonstrate that if it is feasible to shield a specific subset of gates from side-channel attacks, then it is possible to conceal all information in a quantum circuit by transpiling it into a new circuit whose depth grows linearly, depending on the quantum computer's architecture. We provide concrete examples of implementing this protection on IBM's quantum computers, utilizing their virtual gates and editing their transpiler.
\end{abstract}


\section{Introduction}

Advances in quantum algorithms and quantum information~\cite{BCK22,MR22,BS16} have created demand for accurate, reliable, and secure quantum computers. Presently, cloud-based vendors such as IBM Quantum~\cite{IBMQ}, Amazon Bracket~\cite{AmzB}, and Microsoft Azure~\cite{MicA} allow anyone to run an algorithm on a noisy intermediate-scale quantum (NISQ) device. As more sensitive algorithms and data are entrusted to cloud-based quantum computing services, the supplier would likely offer a contract promising not to use the data for anything other than the task prescribed by the user. Even so, the cloud-based computing service needs to be careful about the side-channel information leaked when running a quantum algorithm on the data. In general, side-channel information can be thought of as the information leaked via any physical interactions the computer has with its environment. Particularly in the context of cryptography, side-channel attacks have been executed using the acoustics, electromagnetic radiation, timing, and power consumption of the computer. As shown in \cite{XES23}, if quantum algorithms are implemented in their na\"{i}ve way, then a power-based side-channel attack can usually identify the algorithm being run and even reconstruct most of the algorithm itself. This poses a problem for the user and the cloud-based quantum computing suppliers if the power drawn by the computer is readable from an untrusted third party. 

In this paper, we provide information-theoretic tools and demonstrate methods to enhance protection against side-channel attacks on quantum computers. These tools are crafted exclusively for quantum computers and the quantum circuit model, meaning they are not easily transferable to classical computers. We illustrate that virtual gates, a specific subset of gates used by \cite{IBMQ}, provide resistance to side-channel attacks through our devised technique of ``virtual gate masking.'' In~\cite{XES23}, the authors mention the possibility of substituting real gates with virtual gates. In our paper, we make this concrete by developing a transpilation process that masks all gates in the circuit with virtual realizations, while also concealing information about gate placement.

There is a regime called ``blind quantum computation'' ~\cite{BFK09} that provides an information-theoretic secure way for a client to perform a quantum computation without the cloud computing service even knowing what algorithm is being run. However, these protocols require the existence of a quantum channel between the client and server, as well as a device from the client that has the ability to create single qubit quantum states. Instead, we consider a scenario in which the client does not have a quantum device, trusts the cloud provider, but requires protection from adversaries in proximity to the quantum computer.

In this work, we discuss 
quantum side-channels and propose a modification to the transpilation process that masks the resulting gate layout without changing the algorithmic behavior of the circuit. We start with an overview of quantum side-channel attacks and define various types of information that a side-channel attack might aim to extract from a quantum algorithm in~\Cref{sec:QSCA}. In~\Cref{sec:MaskingTranspiler}, we outline a masking process that enables us to conceal all the information in the quantum algorithm, assuming a certain subset of gates cannot be identified. In~\Cref{sec:VirtualGates}, we discuss the concept of a virtual gates, which are used on IBM's superconducting quantum computers. Finally, in~\Cref{sec:VGM}, we apply our new machinery outlined in~\Cref{sec:MaskingTranspiler} to provide a new transpilation process, allowing IBM's quantum computers to achieve information-theoretic security against side-channel attacks, assuming that virtual gate information is undetectable. 
 We leave an implementation of our masking process and experimental verification to future work.

\section{Quantum Side-Channel Attacks}
\label{sec:QSCA}

Side-channel attacks on any type of computer depend heavily on the computer's hardware and architecture. This makes side-channel attacks against quantum computers particularly tricky in the age of NISQ devices since their hardware and architecture are constantly evolving \cite{LJV23}. Our focus is not on how threat models will be physically realized---instead, we examine the types of information a quantum side-channel attacker.

\begin{figure}[htbp]
    \centering
    \includegraphics[width=0.75\columnwidth]{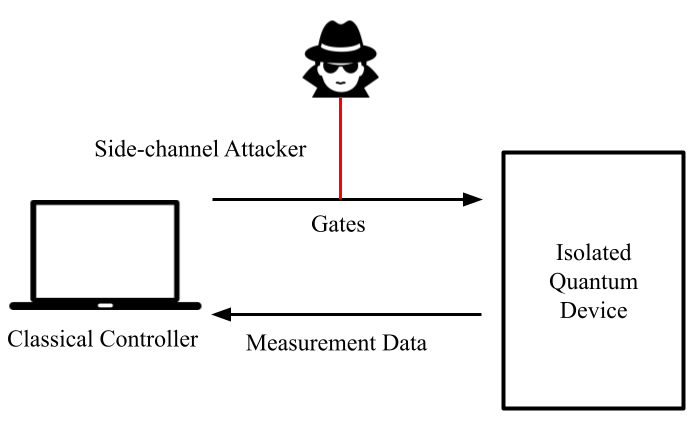}
    \caption{Our side-channel attack model.}
    \label{fig:sidechanneldiagram}
\end{figure}

Unlike classical side-channel attacks, which attempt to recover data used in an algorithm, the ultimate goal of a quantum side-channel attack is \emph{circuit reconstruction} \cite{XES23}. Using side-channel information to directly recover the computer's qubits, rather than its evolution, would effectively measure parts of its internal quantum state, a process limited by the  \emph{no-cloning theorem} which prohibits creating identical copies of an unknown quantum state. This would not only affect the computation of the quantum computer but also suggest that the system's decoherence is significant and unfit for quantum computation. This gives quantum computers a unique advantage over classical computers when defending against side-channel attacks. Any attack which reveals (partial information about) the quantum state means the state will (partially) collapse, impacting the output of the algorithm possibly in a detectable manner. In many current quantum computer realizations, quantum algorithms start 
with $n$ fresh qubits, each in the state $\ket{0}$. 
These qubits evolve over time to a state which is then measured~\cite{MWSCG17}.  
The measurement data is information a side-channel attacker could try to obtain, but this information is purely classical, i.e., not quantum, after the measurement is made. We do not explore side-channel attacks targeting measurement data in this work; rather, we focus on attacks that attempt to learn something about the algorithm itself. 

A quantum algorithm can be parameterized by the sequence of gates used to evolve its qubits, which can be written as a quantum circuit. While there are other models for quantum computers, such as the measurement-based model~\cite{BBD09}, today's largest and most used quantum computers use superconducting qubits and the circuit model~\cite{BDGGN22}. The goal of circuit reconstruction is to identify this sequence of gates. The implementation of a quantum algorithm has details relevant to side-channel attacks that are not needed for theoretical quantum algorithms. Our goal here is to outline the information contained in a physically implemented quantum circuit.

A \emph{universal gate set} is a set $S$ of gates, represented by unitary matrices, that form a generating set for all gates under multiplication and the tensor product. In practical terms, quantum computers apply only the gates in their universal gate set when physically executing a quantum algorithm. For our definition of a universal gate set, we assume, without loss of generality, that it always contains $I_2$, the $2\times 2$ identity unitary. This is the gate which does nothing, i.e., it leaves the qubits alone.
  
\begin{definition}
    An $n$-qubit \textit{quantum circuit} is described by a tuple $C = (U_1, \dots, U_T)$ where each $U_i$ is a unitary  on $n$ qubits with a fixed tensor decomposition $U_{i} = \bigotimes_{j = 1}^\ell V_{i,j}$. Each $U_{i}$ is called a \textit{time step unitary} and each $V_{i,j}$ is a unitary.
    An $n$-qubit \emph{implemented quantum circuit} using a universal gate set $S$ is a quantum circuit where each $V_{i,j}$ is a gate in $S$ that can physically be executed by the quantum computer.

\end{definition} 

We usually denote a quantum circuit by $C$ and express it via its time step unitaries $(U_1, \ldots, U_T)$. There may be many ways to realize a quantum algorithm as a quantum circuit, but when we refer to a quantum circuit $C$, we assume a choice for $(U_{1}, \dots, U_{T})$ has been made along with a tensor decomposition for each $U_{i}$. When a quantum circuit is implemented, we mean the gates in this decomposition are in $S$. The integer $T$ is the circuit's \emph{depth}. 

\begin{definition}
    A gate $V$ in a circuit $C$ has four attributes:
    \begin{enumerate}\itemsep-0.0ex
        \item An \emph{identification} (or \emph{label}): a unitary matrix that encodes the operation that \(V\) performs on the qubits it acts on;
        \item An \emph{index}: the time step at which it is applied in $C$. Formally, the index is an integer $ind \in \{1, \dots, T\}$ such that $V$ is a gate of the time step unitary at index $ind$;
        \item A \emph{wire label}: a list of qubits the gate is applied to. We label the $n$ qubits $\{1, \dots, n\}$ and let the wire label be the subset of qubits on which the gate $V$ acts; and,
        \item A \emph{size}: the number of qubits $V$ acts on.
    \end{enumerate}
\end{definition}
In our paper, when we refer to a gate $V$, we do not only refer to its matrix label, we also imply it is in a quantum circuit with a fixed position. We implicitly assume that $V$ is in the tensor decomposition of some time step unitary $U_{i}$ when we say $V$ is (used) in $C$. Note that a gate's size is included in its wire label but its size does not determine its wire label. If a gate has the identification of an identity matrix of any size, we call it an \textit{identity gate}.

\begin{example} 
\label{ex:weirdCircuits}
    When specifying a circuit, one must be careful with the tensor decomposition of its time step unitaries. Consider the circuits $C_1 = (U_{1}) = (I \otimes H)$ and $C_2 = (U'_{1}) = ((I \otimes H))$ pictured in~\Cref{fig:identity}. Note that $C_1$ has two gates: one with identification $I$, index 1, and wire label $1$, and one with identification $H$, index 1, and wire label $2$. However $C_2$ has one gate with identification $I \otimes H$, index 1, and wire label $\{1, 2\}$. Both these circuits are functionally equivalent, but are technically different circuits.
\end{example}
    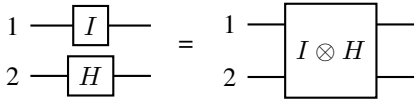
\begin{figure}[ht!]
    \vspace{-12pt}
	\centering
	\begin{quantikz}
            \lstick{1}& \gate{I} &\qw \\[\cond]
            \lstick{2}& \gate{H} &\qw
        \end{quantikz} 
        \hspace{0.05in}
        =
        \hspace{0.0125in}
        \begin{quantikz}
            \lstick{1}& \gate[wires=2]{I \otimes H} &\qw  \\[\cond]
            \lstick{2}& &\qw 
        \end{quantikz}
	\caption{Pictured is $C_1$ on the left and $C_2$ on the right from~\Cref{ex:weirdCircuits}. These circuits do the same thing, but have different representations in terms of gates.
 }
	\label{fig:identity}
    \end{figure}

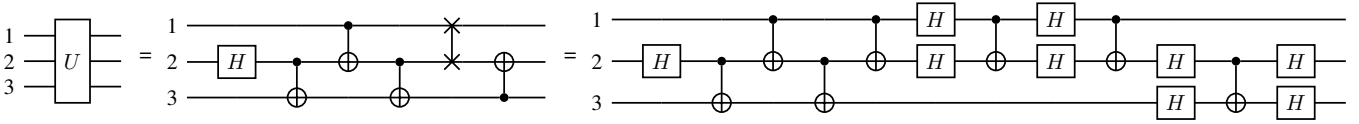
\begin{figure*}[ht] 
	\centering
 \resizebox{\textwidth}{!}{
	\begin{quantikz}
            \lstick{1}& \gate[wires=3]{U} &\qw & \\[\cond]
            \lstick{2}& &\qw &\\[\cond]
            \lstick{3}& &\qw &
        \end{quantikz}\hspace{-0.175in} =
        \begin{quantikz}
            \lstick{1}&\qw &\qw &\ctrl{1} &\qw &\swap{1} &\qw &\qw  \\[\cond]
            \lstick{2}&\gate{H} &\ctrl{1} &\targ{} &\ctrl{1} &\targX{} &\targ{} &\qw\\[\cond]
            \lstick{3}&\qw &\targ{} &\qw &\targ{} &\qw &\ctrl{-1} &\qw
        \end{quantikz}\hspace{0.025in} =
	\begin{quantikz}
            \lstick{1}&\qw &\qw &\ctrl{1} & \qw &\ctrl{1} &\gate{H} &\ctrl{1} &\gate{H} &\ctrl{1} &\qw &\qw &\qw &\qw  \\[\cond]
            \lstick{2}&\gate{H} &\ctrl{1} & \targ{} & \ctrl{1} &\targ{} &\gate{H} &\targ{} &\gate{H} &\targ{} &\gate{H} &\ctrl{1} &\gate{H} &\qw\\[\cond]
            \lstick{3}&\qw &\targ{} &\qw &\targ{} &\qw &\qw &\qw &\qw &\qw &\gate{H} &\targ{} &\gate{H} &\qw
        \end{quantikz}
        }
	\caption{Different circuits for the quantum teleportation algorithm. 
  The circuit $C_1$ has a single gate with identification $U$, index 1, wire label $\{1, 2 , 3\}$, and size 3; the circuit 
  $C_2 = (U_{1}, \dots, U_{6})$ where $U_{1} = I \otimes H \otimes I$, $U_{2} = I \otimes \textrm{CNOT}$, and so on. The first Hadamard gate on the left of $C_2$ has identification $H$, index 1, wire label 2, and size 1. In $C_3$, the last two gates have identification $H$, index 12, size 1, and wire labels 2 and 3 respectively.}
	\label{fig:eqcircs}
\end{figure*}

\begin{example}
    Consider the three circuits in~\Cref{fig:eqcircs} (on the next page). 
     The first circuit is $C_1 = (U_{1}) = (U)$, where $U$ is a unitary matrix. 
    The second circuit $C_2$ exhibits a decomposition of the gate $U$ in terms of ``simpler'' gates more commonly used. 
    The third circuit $C_3$ is an implemented quantum circuit written in the universal gate set $S = \{I, \textrm{CNOT}, H, X_{\pi / 2}, T\}$. 
    The transformation from $C_1$ to $C_3$ or from $C_2$ to $C_3$ is an example of \emph{transpilation}, a process that takes an arbitrary gate as input and implements it using a fixed gate set. 
    
    Note for all of our circuits, we have fixed an ordered labeling of the qubits. 
    A frequently used two-qubit gate is the CNOT gate, which has a control qubit and a target qubit. 
    If you switch the control with the target, you get a completely different gate even though they act on the same two qubits. 
    Thus, these two gates are different and require different matrix identifications even though they are both commonly referred to as CNOT. 
\end{example}

We now consider circuit reconstruction. 

\begin{definition} Let \(C\) be a quantum circuit.
    \begin{enumerate}\itemsep-0.0ex
        \item The \textit{total information} of $C$ is the identification, index, and wire label of the gates in $C$.
        \item For a subset $R \subset S$, the $R$-\textit{absent information} of $C$ is the identification, index, and wire label of all gates not identified by an element of $R$.
        \item The \textit{positional information} of $C$ is the index and wire label information of its gates.
        \item For a subset $R \subseteq S$, the $R$-\textit{positional information} is the $R$-\textit{absent information} of $C$ together with the index and wire labels of gates in $R$.
 
    \end{enumerate}
\end{definition}

The ideal side-channel attack would be able to recover the total information of $C$. The side-channel attacks studied in \cite{XES23} use power consumption and timing information in their attacks. The strongest attack they implement is the \emph{Per-Channel Power Single Trace Attack}, which allows the attacker to measure the power traces of the drive and control channels in the quantum computer separately---in this paper, we assume this powerful threat model, which is described fully in~\cite[Section 3.2.5]{XES23}. In our formalism, this translates to knowledge of the $R$-absent information of the circuit for a specific subset of gates called virtual gates which we discuss in~\cref{sec:VirtualGates}.

\section{Masking in the Transpiler}
\label{sec:MaskingTranspiler}

\subsection{Masking certain gates} 

The term \textit{transpilation} is used by Qiskit \cite{Qiskit} to refer to the process of transforming a high-level description of a quantum algorithm into a circuit using gates that can be physically executed by the quantum computer. One can think of the transpiler as the procedure that decomposes a quantum circuit $C$ into an implemented quantum circuit $B = (U_{1}, \dots, U_{n})$ where each gate of $B$ is in the universal gate set $S$. In the following, we outline a modification to the transpiler that gives extra protection against circuit reconstruction under the assumption that a certain subset of gates $R \subseteq S$ are difficult to detect. From here on out, we assume the quantum computer operates on $n$ qubits and we let $N$ be the maximum size of a gate at the time our transpiler modification occurs.

\begin{definition}
    \label{def:CoveringSet}
    A subset \(R\) of a universal gate set $S$ is called a \textit{covering gate set} if for each $m\in\{1, \ldots, N\}$, there is a fixed ordered tuple of gates in $S$, $(S_{1}, \dots, S_{r_{m}})$, where each $S_{i}$ acts on $m$ qubits and is a tensor product of gates in $S$. These $m$-qubit gates must satisfy that any $m$-qubit gate $U$ can be decomposed as 
    \begin{equation}
        \label{eq:CoveringSet}
        U = \prod_{i = 1}^{r_{m}} R_{i}S_{i}
    \end{equation} 
    for some choice of $(R_{1}, \dots, R_{r_{m}})$ where each $R_{i}$ is a tensor product of gates in $R$.
\end{definition}
\begin{procedure}
    \label{proc:RGateMask}
    Given a universal gate set $S$ and a covering gate set $R \subseteq S$, we define $R$-\textit{gate masking} to be the following modification to a deterministic transpilation process:
    \begin{enumerate}
        \item For each $m$ up to $N$, fix an ordered tuple $(S_{1}, \dots, S_{r_{m}})$ given by~\Cref{def:CoveringSet}.
        \item Given a high-level quantum circuit $C$, decompose each non-identity gate $V$ in $C$ using \Cref{eq:CoveringSet} with $m$ equal to the number of qubits $V$ acts on.
	\item Proceed with transpilation as usual.
    \end{enumerate}
\end{procedure}
\begin{remark}
Step (1) serves as a preprocessing step, performed prior to receiving an algorithm for transpilation. Step (3) is necessarily vague due to the dependence of hardware and architecture on transpilation, which involves more than just decomposing gates. For example, if the given high-level quantum circuit applies two gates $U$ and $V$ at the same time step, the hardware of the computer may require $U$ and $V$ to act on the same number of qubits. Moreover, the application of $U$ and $V$ may be feasible only when they act on qubits physically separated from each other within the quantum computer. Because of this, the computer may need to apply $U$ before $V$, or vice versa, resulting in the addition of time steps. We require that the existing transpilation process is deterministic because we want the positional information of a transpiled circuit to be determined by the inputted algorithm.
\end{remark}

\begin{theorem}
    \label{thm:RGateMask}
    Suppose we have a universal gate set $S$, covering set $R$, and two quantum circuits $C$ and $C'$ with the same positional information. If $C$ is transpiled into $B$ and $C'$ is transpiled into $B'$, both with $R$-gate masking, then the $R$-positional information of $B$ and $B'$ are equivalent.
\end{theorem}

\begin{proof} 
Two circuits with equal positional information will have transpiled circuits with equal positional information (noting that we treat $R_i$ as a single gate). This is because each $m$-qubit gate $V$ is transpiled into a sequence of gates containing positional information dependent only on the positional information of $V$ and not its identification. Therefore the positional information of $B$ and $B'$ are equal. Moreover, if we consider the set of gates with identification in $S \setminus R$, they have the same identification in both $B$ and $B'$ since $S_{1}, \dots, S_{r}$ is a sequence of gates fixed by the transpilation process and each gate $V$ gets transpiled as $V = \prod_{i=1}^{r_{m}} R_{i}S_{i}$. Thus, the only difference in $B$ and $B'$ is the identifications of gates in $R$.
\end{proof}

One way to interpret the above theorem is that $R$-gate masking provides a way to hide all identification information in a circuit, given one can hide the identification of the gates in $R$.
Ideally, the set $R$ is large enough so that it is infeasible to simply guess the identification of a gate in an $R$-masked circuit whose identification is in $R$. We show an application where $R$ is the set of virtual gates, which is an infinite set. Note that the sequence $B$ still contains the positional information of $C$, even if the identification of gates in $R$ are unknown. For example, if $C$ first applies a single-qubit gate to wire 1, then a two-qubit gate to wires $\{2, 3\}$, an attacker with $R$-positional information will still be able to recover this information from $B$. Luckily, there are ways to mask these properties as well.

\subsection{Masking gate positions} 
\label{sec:MaskingGatePos}

We now provide a procedure to mask the positional information of any quantum circuit. This is achieved by creating a subcircuit that applies a gate to every possible wire label.

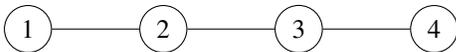
\begin{figure}[ht!]  
    \centering
    \begin{tikzpicture}[scale=0.9]
        \node[unfilled] at (-3, 0) (v_1) {1};
        \node[unfilled] at (-1, 0) (v_2) {2};
        \node[unfilled] at (1, 0) (v_3) {3};
        \node[unfilled] at (3, 0) (v_4) {4}; 
        \draw (v_1) -- (v_2);
        \draw (v_2) -- (v_3);
        \draw (v_3) -- (v_4); 
    \end{tikzpicture}
    \caption{A possible layout of $4$ physical qubits, called a 1D array. If two qubits can have a gate applied to them, they are connected with an edge. These are called \emph{nearest neighbor interactions}.} 
    \label{fig:imbLagos0}
\end{figure}

We first discuss the possible wire labels in a circuit. In a high-level description of a quantum circuit, any two-qubits can share a gate.
At the physical hardware level, this is not always the case. 
In current superconducting quantum computers, there is a notion of a \emph{topology} which refers to the physical layout of the qubits. See Figure~\ref{fig:imbLagos0} as an example. The topology imposes a restriction on which multi-qubit gates can be applied.

Let $W$ be the set of all possible wire labels up to size $N$ that the topology of the quantum computer allows to be applied. Note that sets in $W$ are subsets of qubit indices that represent multi and single qubit gates based on the topology.

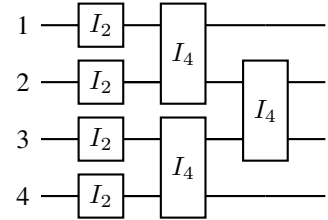
\begin{figure}[ht!]
	\centering
	\begin{quantikz}
\lstick{1} & \gate{I_2} & \gate[wires=2]{I_4} & \qw & \qw \\[-9pt]
\lstick{2}& \gate{I_2} & \qw & \gate[wires=2]{I_4} & \qw \\[-9pt]
\lstick{3}& \gate{I_2} & \gate[wires=2]{I_4} & \qw & \qw \\[-9pt]
\lstick{4}& \gate{I_2} & \qw & \qw & \qw 
	\end{quantikz}
	\caption{Using the topology in Figure~\ref{fig:imbLagos0}, assuming we can apply only single-qubit gates and two-qubit gates, we have $N = 2$ and possible wire-labels $W = \{1\}, \{2\}, \{3\}, \{4\}$, $\{ 1,2 \}$, $\{ 2,3 \}$, and $\{ 3,4 \}$. The circuit $Y$ with minimum depth is shown above with depth $p = 3$.}
	\label{fig:DefineY}
    \end{figure}

\noindent We now define a circuit which may seem odd at first, but it will be a building block to mask positional data. Let $Y$ be a circuit consisting solely of identity gates in the following manner. 
The circuit $Y$ has an identity gate that acts on qubits with indices in $w$ for every ${w \in W}$. Since the construction of $Y$ is not unique, let us fix one with minimum depth, which we denote as $p$. See Figure~\ref{fig:DefineY} for an example of such a \(Y\).

\begin{figure}[htbp!]
	\centering
\resizebox{0.8\linewidth}{!}{
	\begin{quantikz}
\lstick{1}	    &\gate[wires=2]{V_{i,1}} & \qw \\[\cond] 
	   \lstick{2} &\qw  &\qw \\[\cond]
	    \lstick{3}&\gate{V_{i,2}} &\qw \\[\cond]
	    \lstick{4}&\gate{V_{i,3}} &\qw  
	\end{quantikz}\hspace{0.05in}=\hspace{0.025in}
		\begin{quantikz}
\lstick{1}& \gate{I_2} & \gate[wires=2]{V_{i,1}} & \qw & \qw \\[\cond]
\lstick{2}& \gate{I_2} & \qw & \gate[wires=2]{I_4} & \qw \\[\cond]
\lstick{3}& \gate{V_{i,2}} & \gate[wires=2]{I_4} & \qw & \qw \\[\cond]
\lstick{4}& \gate{V_{i,3}} & \qw & \qw & \qw 
	\end{quantikz}
 }
	\caption{Using the topology in Figure~\ref{fig:imbLagos0}, we can replace the time step unitary $U_i \coloneqq \otimes_{j=1}^\ell V_{i,j}$ on the left-hand side with the functionally equivalent unitary $Y_i$ on the right-hand side. Note that the positional data of $Y_i$ is the same as in $Y$ as seen in~\Cref{fig:DefineY}, regardless of how $U_i$ is defined.}
	\label{fig:DefineYi}
    \end{figure}
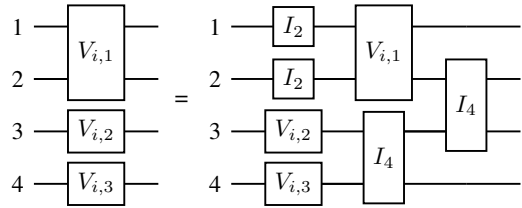

Now that we have fixed this circuit $Y$, we show how any time step unitary $U_i$ can be written to have the exact same positional data as $Y$. This is achieved by replacing the identity gates in $Y$ with $V_{i,j}$ gates in $U_i$, forming $Y_i$. Then, we $R$-gate mask both the identity and $V_{i,j}$ gates in $Y_i$, effectively hiding the positional information of $U_i$. We give an example of this in~\Cref{fig:DefineYi}. 

\begin{procedure}
    \label{proc:totalRGateMask}
    Fix a hardware topology, a  universal gate set $S$, and a covering gate set $R \subseteq S$. Define \textit{total} $R$-\textit{gate masking} to be the following modification to a transpilation process: given a circuit $C = (U_1, \ldots, U_T)$ such that for each time step unitary $U_i = \otimes_{j=1}^\ell V_{i,j}$, the gate $V_{i,j}$ acts on at most $N$ gates:  
    \begin{enumerate}
    \item 
    Define the new circuit $B$ consisting of $T$ consecutive instances of the circuit $Y$. Let $Y_i$ denote the subcircuit of $B$ corresponding to the $i$th instance of $Y$ in $B$. 
    \item For each time step unitary $U_i = \otimes_{j=1}^\ell V_{i,j}$, each $V_{i,j}$ acts on qubits with indices equal to some $w \in W$. For each $V_{i, j}$, replace the identity gate in $Y_i$ that acts on $w$ with $V_{i,j}$.
    \item Perform $R$-gate masking on $B$ but also mask all the identity gates, meaning we remove the restriction pertaining to identity gates in step (2) of~\Cref{proc:RGateMask}.
    \end{enumerate}
\end{procedure}

\begin{remark} In step (3), masking does not necessarily need to be applied to all identity gates, only the set of gates that correspond to $w \in W$. Finally, note that forming $Y$ can be done in preprocessing, much like step (1) of $R$-gate masking. 
\end{remark} 

\begin{example}
    Assume again we have $N = 2$ and the topology in~\Cref{fig:imbLagos0}.
    \begin{figure*}[htbp!]
	\centering
 \resizebox{0.85\textwidth}{!}{
	\begin{quantikz}
\lstick{1}	    &\gate[wires=2]{U_1} &\qw &\qw\\[\cond]
\lstick{2}	    &\qw &\gate{U_3} &\qw\\[\cond]
\lstick{3}	    &\gate{U_2} &\qw &\qw\\[\cond]
\lstick{4}	    &\qw &\gate{U_4} &\qw
	\end{quantikz}\hspace{0.075in} =
	\begin{quantikz}
\lstick{1}	    &\gate{I} &\gate[wires=2]{U_1} &\qw &\qw &\gate{I} &\gate[wires=2]{I \otimes I} &\qw &\qw  &\qw\\[\cond]
\lstick{2}            &\gate{I} &\qw &\gate[wires=2]{I \otimes I} &\qw &\gate{U_3} &\qw &\gate[wires=2]{I \otimes I} &\qw &\qw\\[\cond]
\lstick{3}            &\gate{U_{2}} &\gate[wires=2]{I \otimes I} &\qw &\qw &\gate{I}  &\gate[wires=2]{I \otimes I} &\qw &\qw &\qw\\[\cond]
\lstick{4}	    &\gate{I} &\qw &\qw &\qw &\gate{U_4} &\qw &\qw &\qw &\qw
	\end{quantikz}
 }
	\caption{An example of total $R$-gate masking.}
	\label{fig:totalRGateMaskex}
    \end{figure*}
In~\Cref{fig:totalRGateMaskex}, let the quantum circuit on the left with depth two be $C$. We have already formed our desired $Y$ time step unitary in~\Cref{fig:DefineY}. We create $B$ from 2 copies of $Y$ and then for each $i$ from $1$ to $T$, we insert the non-identity gates from in $U_i$ into $B$ by replacing the identity gates in $Y_i$ with the same wire label. Note in~\Cref{fig:totalRGateMaskex}, the identity gates in $Y_{1}$ and $Y_{2}$ are only shown if they uniquely correspond to a wire label $w \in W$. Only the identity gates that are shown need to be masked in total $R$-gate masking.
\end{example}
\begin{theorem}
    \label{thm:totalRGateMask}
    Suppose we have a universal gate set $S$, a covering set $R$, a fixed topology, and two quantum circuits $C$ and $C'$ of the same depth with gates up to size $N$.
    If $C$ is transpiled into the circuit $B$ and $C'$ is transpiled into the circuit $B'$, both using total $R$-gate masking, then the $R$-positional information of $B$ and $B'$ are equivalent.
\end{theorem}
\begin{remark} 
Note that we can always make the circuits $C$ and $C'$ have the same depth by adding in identity time step unitaries.
\end{remark}
\begin{proof}
Since both circuits have gates of size at most $N$ and the same depth,  the intermediate circuits $D$ and $D'$ produced at the end of step (2) of Procedure~\ref{proc:totalRGateMask} on input $C$ and on input $C'$ will have the same positional information as the circuit consisting of $T$ consecutive instances of the circuit $Y$. Therefore the positional information of $D$ and $D'$ are equal.  
By Theorem~\ref{thm:RGateMask}, the $R$-positional information of $B$ and $B'$ are equal since the positional information of $D$ and $D'$ are equal.  

\end{proof}
\begin{remark}
    Notice that $C$ and $C'$ no longer have to share the same positional information like they do in~\Cref{thm:RGateMask}. We have successfully hidden the positional information of a circuit given the identification of gates in $R$ are undetectable.
\end{remark}

\subsection{The depth of masked circuits} 
\label{sec:YDepth}

Now, we discuss the overhead incurred by these transpilation processes in terms of circuit depth. 

\begin{theorem}
    Suppose we have a quantum algorithm $\mathcal A$, a universal gate set $S$, a covering set $R$, and a fixed topology. 
    Suppose further that there is an existing transpiler that transpiles $\mathcal A$ into the quantum circuit $C$ with depth $T$ after considering the topology and universal gate set $S$. Let $B$ be the circuit of depth $T_B$ transpiled by the same transpiler with $R$-gate masking and let $D$ be the circuit of depth $T_D$ transpiled using total $R$-gate masking. Let $Y$ be the circuit of depth $p$ as defined in the previous section. Set $r = \max_{1\leq m \leq N} \{r_{m}\}$ as defined in~\Cref{proc:RGateMask}. Then 
    \(
	T_B \le 2rT \mbox{, and } 
    T_D \le 2rpT.
    \)
\end{theorem} 

\begin{remark} 
Additional physical implementation restrictions may need to be considered by the computer when transpiling, adding depth to the circuit. For example, a quantum computer may only allow a single gate to be applied in every time step, increasing the depth by a factor of at most $n$ for every time step in the original circuit. Since the circuit $Y$ does not account for these physical restrictions, we assume the transpiler addresses them after our masking process has been executed. See~\Cref{sec:VGM} for an example. 
\end{remark} 

We now consider the depth of the circuit $Y$, detailed in the last subsection. In superconducting quantum computers, it is common to represent their topology with a graph $G = (V, E)$ where each vertex is the index of a qubit after an ordering of the qubits has been fixed. An edge between qubits $u$ and $v$ means that the computer can apply a two-qubit gate to $u$ and $v$. In this case, we assume that the topology only allows for one and two-qubit interactions, meaning $N = 2$ as we cannot apply gates to three or more qubits.
Indeed, many universal gate sets contain only $1$ and $2$-qubit gates.   

The first requirement of the circuit $Y$ is that it must apply a single qubit identity gate to all qubits. Regardless of the topology, we can always apply a single qubit gate to every qubit in one time step.

To discuss the minimum number of time steps required to apply an identity gate to all possible two-qubit pairs, we need to recall the edge chromatic number of a graph $G$. The  {\em edge chromatic number}, denoted $\chi'(G)$, is the minimum number of colors required to color the edges of $G$ so that if two edges share a common vertex, they have different colors. We can trivially color any graph with $|E|$ colors, but $\chi'(G)$ it may be that $
\chi'(G) < |E|$. 

Suppose two edges are colored ``blue'' in a valid edge coloring. This corresponds to two distinct $2$-qubit gates with no qubits in common, meaning they can be applied in the same time step in the circuit $Y$. In fact, all edges corresponding to the same color can be applied in parallel and therefore we can apply all possible two-qubit identity gates in at most $\chi'(G)$ time steps.

\begin{theorem}
    Suppose $N = 2$ and the topology of the quantum computer is given by a graph $G$. Let $Y$ be the minimum depth circuit that implements an identity gate for every possible wire label $w \in W$. Then the depth of $Y$ is either $\chi'(G)$ or $\chi'(G) + 1$.
\end{theorem} 

\begin{proof} 
The single-qubit gates in $Y$ can be implemented in one time step and the two-qubit gates in $Y$ with $\chi'(G)$ time steps. Now, we argue that $\chi'(G)$ is the minimum number of time steps we can implement the two qubit identity gates in $Y$. Suppose there exists a circuit $C = (U_1, \ldots, U_d)$ of depth $d < \chi'(G)$ that implements a two qubit identity gate for every edge $e \in E$. We will color the graph with $d$ colors, leading to a contradiction. For every two-qubit gate in $U_1$, color the corresponding edge ``color 1''. For every two-qubit gate in $U_2$, color the corresponding edge ``color 2,'' and so on. Doing this, we would eventually color the whole graph with $d < \chi'(G)$ colors, as desired. So we do in fact need at least $\chi'(G)$ many time steps in a circuit $C$ implementing all of the two-qubit gates. Since \(Y\) requires at most one more time step to accommodate all single qubit gates, we see that the depth of \(Y\) is either \(\chi'(G)\) or \(\chi'(G) + 1\).\qedhere
\end{proof}  

We have now bounded the depth of $Y$ in terms of the edge chromatic number of $G$. Vizing's Theorem~\cite{vizing1965critical}  states that $\Delta(G) \leq \chi'(G) \leq \Delta(G) + 1$, where \(\Delta(G) \leq n-1\) is the maximum degree of any vertex in \(G\). This yields Corollary~\ref{cor:delta}. 

\begin{corollary}
\label{cor:delta}
    If the gates in $S$ act on at most $N=2$ qubits, the depth of $Y$ is at most $\Delta(G) + 2 \le n + 1$, where $n$ is the number of qubits.
\end{corollary}

\section{Virtual Gates}
\label{sec:VirtualGates}
A qubit can be interpreted non-uniquely as a point on a three-dimensional unit sphere and single-qubit unitaries as (certain) maps from the unit sphere to itself. 
Single-qubit unitary transformations can be decomposed as rotations around the $x$, $y$, and $z$ axes of the sphere. These rotations are denoted $X_\theta$, $Y_\theta$, and $Z_\theta$ respectively where $\theta$ denotes the angle of rotation about that axis.

We can write any single qubit unitary as
\begin{align}
\nonumber
    U(\theta, \phi, \lambda) &\coloneqq Z_\theta X_\phi Z_\lambda\\
    &= Z_{\phi - \pi/2} X_{\pi/2} Z_{\pi - \theta} X_{\pi/2} Z_{\lambda - \pi / 2}
    \label{virtualdecomp}
\end{align} 
for some choice of angles $\theta$, $\phi$, and $\lambda$. 
This means that the gate set $\{X_{\pi/2}$, $Z_\theta\}_{\theta \in [0, 2\pi)}$ is universal for single-qubit gates~\cite{MWSCG17}.

We discuss how such a single qubit unitary $U(\theta,\phi,\lambda)$ is applied on IBM's superconducting quantum computer. On their quantum computer, gates are applied to qubits using pulses that consist of an amplitude $\Omega(t)$, a frequency $\omega$, and a phase $\gamma$. The pulse is a microwave generated by an arbitrary waveform generator (AWG), a constant amplitude microwave generator, and an IQ mixer \cite{XES23}. The collection of these machines allows one to create an arbitrary microwave of the form $\Omega(t)\cos(\omega t - \gamma)$. The AWG is the component that is classically programmable and used for shaping the wave \cite{MWSCG17}. Assuming a constant amplitude pulse $\Omega$ for duration $T$, the unitary applied by the pulse is
\begin{equation*} 
A(\gamma,\Omega,T) \coloneqq e^{-i\frac{\Omega T}{2}\bp{\cos(\gamma)X_\pi + \sin(\gamma)Y_\pi}}. 
\end{equation*} 
The unitary $A(\gamma,\Omega,T)$ is a rotation around some axis.
The key point regarding the virtual gate is that adjusting $\gamma$ rotates the axis of rotation of $A$ about the $z$-axis. For example, if $\gamma = 0$, the unitary $A(0,\Omega,T)$ is a rotation by $\Omega T$ around the $x$ axis; likewise if $\gamma = \pi / 2$, the unitary $A(\pi / 2, \Omega, T)$ is a  rotation by $\Omega T$ around the $y$ axis \cite{MWSCG17}. This means the unitary $U(\theta, \phi, \lambda) = Z_\theta X_\phi Z_\lambda$ 
can be applied by adjusting the phase for the pulse of $X_\phi$ by $\theta$ and adjusting the phase of all future gates by $\lambda$. Also, since measurements yield the classical bits $0$ or $1$ with a probability given by a function of the position along the $z$ axis,
a rotation about the $z$ axis of the qubit has no effect on the measurement probabilities.

The physical gate set employed by IBM is composed of four types of gates: $Z_\theta, X_{\pi/2}$, $X$, and CNOT \cite{JAA22}. This gate set was chosen due to the efficiency of using virtual gates which are given by $Z_\theta$ for some angle $\theta$.
To minimize error, sending as few pulses as possible to the qubits is ideal. In this paper, we demonstrate the two-fold usefulness of these virtual gates: they not only provide an efficient implementation of gates but also offer resistance to side-channel attacks through our process of \textit{virtual gate masking} (discussed in the next subsection). In IBM's universal gate set, only $X_{\pi / 2}, X$, and CNOT are able to be detected using side-channel attacks on the pulses of the quantum computer, unless one is able to externally detect the phase of a pulse sent to a qubit. There is certainly a possibility that side-channel attacks with direct access to the AWG or the classical controller could detect the use of a $Z_\theta$ gate, as well as $\theta$. In this context, virtual gate masking has the effect of concentrating the side-channel vulnerabilities of the quantum computer to classical devices, which is useful as there is already plenty of literature on protecting against classical side-channel attacks \cite{PPMWB23,LZJZ21}. 
However, the attacks in \cite{XES23} are unable to detect the application of virtual gates using their strongest attempt at a power-based side-channel attack, motivating the following definition.

\begin{definition}
    The \textit{non-virtual information} of a circuit is its $\{Z_{\theta}\}_{\theta \in [0, 2\pi)}$-absent information.
\end{definition}

We show that not being able to detect virtual gates has significant implications for attempts at circuit reconstruction and identification.

\subsection{Virtual gates as covering sets} 
Let us fix the universal gate set to be $S = \{ Z_\theta, X_{\pi/2}, X, CNOT \}$ and $R = \{ Z_\theta \}_{\theta \in [0, 2\pi)}$.
We now prove that $R$ is a covering set when $N = 2$ is the maximum size of a gate that can be applied.

Note that each single-qubit gate can already be decomposed using~\Cref{virtualdecomp}, which gives us our single-qubit sequence for free. Specifically, we have $r_1 = 3$ and $S_1 = X_{\pi / 2}$, $S_2 = X_{\pi / 2}$, and $S_3 = I$.

\begin{figure}[ht!] 
\centering
\resizebox{\columnwidth}{!}{
    \begin{quantikz}
       \lstick{1} &\gate[wires=2]{U} &\qw\\[-9pt]
       \lstick{2} &\qw &\qw
    \end{quantikz}\hspace{0.1in} =
    \begin{quantikz}
        \lstick{1} &\gate{u_1} &\ctrl{1} &\gate{u_2} &\ctrl{1} &\gate{u_3} &\ctrl{1} &\gate{u_4} &\qw\\[-9pt]
        \lstick{2} &\gate{u_5} &\targ{} &\gate{u_6} &\targ{} &\gate{u_7} &\targ{} &\gate{u_8} &\qw 
    \end{quantikz}
    }
    \caption{Decomposition of a two-qubit gate into eight single-qubit gates and three CNOT gates.} 
    \label{fig:virtualtwoqudecomp}
\end{figure}
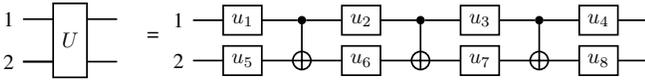

For two-qubit gates, it was proven in~\cite{vidal2004universal} that any two-qubit unitary can be decomposed into eight single-qubit gates and three CNOT gates as shown in Figure~\ref{fig:virtualtwoqudecomp}. 

Since $(\theta_{i}, \phi_{i}, \lambda_{i})$ parameterize $u_{i}$ using~\Cref{virtualdecomp}, we have that $r_2 = 12$ and
\begin{align*}
\allowdisplaybreaks
    S_{1} &= X_{\pi / 2} \otimes X_{\pi / 2} &  S_{7} &= X_{\pi / 2} \otimes X_{\pi / 2}\\
    S_{2} &= X_{\pi / 2} \otimes X_{\pi / 2} & S_{8} &= X_{\pi / 2} \otimes X_{\pi / 2}\\
    S_{3} &= \textrm{CNOT} & S_{9} &= \textrm{CNOT}\\
    S_{4} &= X_{\pi / 2} \otimes X_{\pi / 2} & S_{10} &= X_{\pi / 2} \otimes X_{\pi / 2}\\
    S_{5} &= X_{\pi / 2} \otimes X_{\pi / 2} & S_{12} &= X_{\pi / 2} \otimes X_{\pi / 2}\\
    S_{6} &= \textrm{CNOT} & S_{13} &= I \otimes I.
\end{align*}

\subsection{Virtual Gate Masking}
\label{sec:VGM} 

\begin{figure}[ht!]
    \centering
    \includegraphics[width=\linewidth]{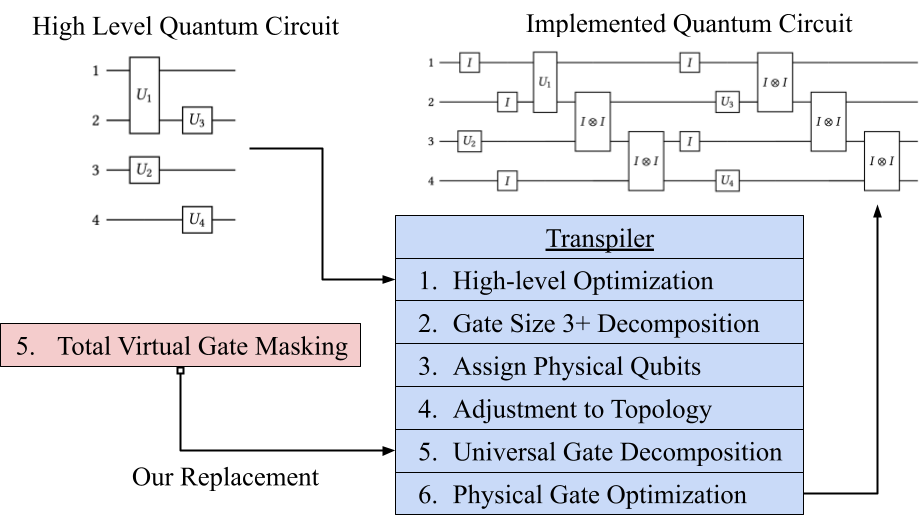}
    \caption{Our proposed modification to IBM's Standard Transpiler on Qiskit.}
    \label{fig:IBMTranspiler}
\end{figure}

Here we discuss using our masking procedure with virtual gates on IBM's quantum computers. 
\begin{definition}
    We call $R$-gate masking \textit{virtual gate masking} when $R$ is the set of virtual gates $\{ Z_\theta \}_{\theta \in [0, 2\pi)}$. 
\end{definition}
Figure~\ref{fig:IBMTranspiler} depicts IBM's Qiskit standard transpiler \cite{QiskitTranspile} and our proposed modification to protect against quantum side-channel attacks. 

The edit we make in both our masking processes is to pass 5, `Translate to Basis Gates,' which is the step that decomposes arbitrary gates into gates in $S$. Thanks to pass 2, we can assume that each gate is at most size two in the stage, meaning that $N = 2$ is the maximum gate size in the circuits we are considering during our masking process.
The preprocessing needed for virtual gate masking has been done in the previous subsection. Pass 5, `Translate to Basis Gates,' will be modified so that size 1 gates are decomposed using~\Cref{virtualdecomp} and size 2 gates are decomposed using the equivalence depicted in~\Cref{fig:virtualtwoqudecomp}, with the single-qubit gates $u_{i}$ still being decomposed with~\Cref{virtualdecomp}. Note that when a size 1 or 2 gate is applied, an attacker that cannot see the identification of a virtual gate will see the sequence $(A, X_{\pi / 2}, A, X_{\pi / 2}, A, I)$ or $(A \otimes A, X_{\pi / 2} \otimes X_{\pi / 2}, A \otimes A, X_{\pi / 2} \otimes X_{\pi / 2}, A \otimes A, A \otimes A, \textrm{CNOT}, \dots)$ for some gates $A$ with unknown identification whenever a gate is applied.

Note that virtual gate masking always translates a single-qubit gate into $6$ gates in $S$, three of which are virtual. One can remove $S_{r_{1}} = I$ to make this 5 instead. A single-qubit gate is always implemented with 5 gates during virtual gate masking, so the extra depth added to the circuit is increased by at most $4$ gates for every single-qubit gate. Similarly, a two-qubit gate is always implemented with a depth of $24$ which can be made into $23$ be removing $S_{r_2} = I \otimes I$. This means the extra depth added to the circuit is at most $23$ for every two-qubit gate.

We need to do additional preprocessing for total virtual gate masking. In particular, we need to create the circuit $Y$ from~\Cref{sec:MaskingGatePos} dependent on the topology of our quantum computer. 
When modifying the transpiler for total virtual gate masking, we add the additional modification to pass 5 outlined in~\Cref{proc:totalRGateMask}. Note in total virtual gate masking, a gate on each wire is masked for every time step in the original circuit so that an attacker blind to the virtual gate identifications does not know which wire label has a gate that is not the identity.

The overhead incurred by total virtual gate masking is discussed in~\Cref{sec:YDepth}. Note that the algorithm that is output in~\Cref{fig:IBMTranspiler} is not the same as the one in~\Cref{fig:totalRGateMaskex}. This is because pass 6 of the transpiler may make some additional modifications to our circuit. For example, in~\Cref{fig:IBMTranspiler} it is assumed that two gates are not applied in the same time step if any of the qubits the gates act on are adjacent to each other in the computer's topology.

\section{Conclusions and Future Work}
Total virtual gate masking provides us with information-theoretic protection against any side-channel attack on a quantum computer, assuming it is possible to hide virtual gate angles from the attacker. Virtual gates are programmed and executed in a classical controller, affecting computation only outside of the classical controller—internally inside the arbitrary waveform generator (AWG) and in the phase of microwave pulses sent to qubits. Assuming that the phase of microwave pulses are physically undetectable, then we effectively reduce side-channel attacks on quantum computers to side-channel attacks on the classical components of the quantum computer. Total virtual gate masking offloads the side-channel security of the quantum computer to just a few classical components, enabling us to use previously well-studied classical methods to protect against side-channel attacks \cite{PPMWB23, LZJZ21}.
Moreover, our procedure could easily be implemented on most current gate-based quantum computing vendors. We explicitly outline how to do so in Qiskit \cite{QiskitTranspile} in~\Cref{sec:VGM}. 
  
At the start of our analysis, we assumed that our quantum computer has an ``isolated quantum device'' that the side-channel attacker is unable to tamper with, because any measurement of the evolving quantum state would introduce error into the quantum computation rendering the computer and the attack useless. That being said, one can envision a very aggressive threat model where an adversary is able to make partial measurements to the quantum state, introducing only small amounts of error into the computation undetected. 
This is a more realistic threat model, where a technical analysis of the trade-off between information gained by the attacker and error introduced to the computation is needed. 
For example, the attacker could attempt to read side-channel information from the 
the measurement channels of the quantum computer. 
It would be interesting to see if these distinct threat models can be combined to access information previously thought secure. 

\bibliographystyle{ACM-Reference-Format}
\bibliography{citation}

\end{document}